# Masses of Isolated Spiral KIG Galaxies, Determined by the Motions of their Faint Companions

V. E. Karachentseva,[1,][*] I. D. Karachentsev,[2] O. V. Melnyk[1]

[1]*Main Astronomical Observatory, National Academy of Sciences of Ukraine, Kyiv, 03143 Ukraine*
[2]*Special Astrophysical Observatory, the Russian Academy of Sciences,
Nizhnij Arkhyz, Karachai-Cherkessian Republic, Russia 369167*

We have updated the classification of late-type galaxies presented in the Catalog of Isolated Galaxies (KIG) using the advanced digital sky surveys. Our search for companions around 959 KIG galaxies revealed 141 neighbors associated with 111 KIG galaxies within the mutual projection separation of less than 330 kpc and the radial velocity difference not exceeding 500 km s$^{-1}$. Typical luminosity of the companions turned out to be weaker than the luminosity of the main galaxies by more than an order of magnitude. Considering these small companions as test particles that move around the KIG galaxies along the Keplerian orbits with eccentricity of $e \simeq 0.7$, we estimated the total (orbital) masses of spiral KIG galaxies. Their average orbital mass-to-$K$-band luminosity ratio, $(20.9 \pm 3.1) M_\odot/L_\odot$, is in a good agreement with the corresponding value for the nearby Milky Way, M 31 and M 81-type massive spirals. Isolated disk-shaped galaxies have an on the average 2-3 times smaller total-mass-to-stellar-mass ratio than those of isolated bulge-shaped galaxies.

Ключевые слова: galaxies: isolated—galaxies: late types—galaxies: orbital masses

## 1. INTRODUCTION

The presence of dark matter (DM) in the Universe is now a generally accepted basis for the standard cosmological LCDM model. According to up-to-date ideas, every galaxy is formed inside a dark matter halo (Wechsler and Tinker 2018). However, the physical nature of the dark matter is still unclear since the report by Rubin (1986) at the XIXth IAU General Assemby. We can confidently assert that DM makes up the bulk of the total mass of the Universe (Bahcall 2015, Fukugita and Peebles 2004). Historical and philosophical considerations about the nature of DM are presented in the recent paper by Schombert (2021).

Without touching on the history of the issue and numerous studies on DM in clusters and in galaxy groups, let us mention several papers describing the DM around individual galaxies. The fact that galaxies are surrounded by massive dark halos was for the first time described by Einasto et al. (1974), Frenk and White (1980), White and Rees (1978). According to White and Frenk (1991), dark halos play a fundamental role in the formation and evolution of galaxies.

Let us list here several pioneering observational studies related to this topic. Lynden-Bell and Frenk (1981) proposed to use the kinematics of globular clusters (GC) to determine the total mass of our Galaxy. In their study Lynden-Bell et al. (1983) used radial velocity measurements of carbon stars and planetary nebulae, as well as dwarf companions of our Galaxy and estimated the dynamic mass of the Galaxy within 100 kpc. Based on the observations of rotation curves (RC) in spiral galaxies Rubin et al. (1978, 1980) (and the literature cited there) concluded that flat or increasing RCs indicate the presence of an extended dark halo around these galaxies. Later on, Lapi et al. (2018), Marasco et al. (2021), Posti and Fall (2021) used various RC and GC observational reports to determine the hidden mass around several hundred single galaxies. The authors note that the DM mass calculations were performed both for the late-type (LTG) and early-type galaxies (ETG).

Zaritsky et al. (1993, 1997), Zaritsky and White (1994) searched for faint companions around single spiral galaxies, measured radial velocities of companions and determined the dynamic masses of the central galaxies at the scale of the virial radius of their dark halo. A similar analysis of the kinematics of companions around single galaxies in the Sloan Digital Sky Survey (SDSS DR10, Ahn et al. (2014)) was performed by More et al. (2011), Seo et al. (2020), Wang et al. (2021). The search for faint companions around nearby massive galaxies was continued in Carlsten et al. (2020), Smercina et al. (2018). After determining the distances to the found companions, measuring their radial velocities and exclusion of non-physical neighbors, the remaining companions can be used to calculate the mass of the halo around the central galaxies. Numerous observational, theoretical papers and especially the studies on the numerical modeling of this problem are discussed, for example, in the survey by Wechsler and Tinker (2018).

[*]Electronic address: `valkarach@gmail.com`

Karachentseva et al. (2011) estimated the masses of isolated galaxies from the 2MIG catalog (Karachentseva et al. 2010) based on radial velocities of their small companions. It was found that the median 'mass/$K$-band luminosity' ratio for LTG galaxies (late types) is 3 times lower than that for the ETG galaxies (early types). Karachentseva et al. (2021) calculated the ratio of the total mass $M_T$ to stellar mass $M_*$ for the isolated E and S0 galaxies from the KIG (Karachentseva 1973) catalog using the radial velocity data of companion galaxies. The results for 2MIG and KIG galaxies are in a good mutual agreement. As a continuation of the previous study, here we determine the total mass-to-stellar mass ratio for the isolated spiral galaxies (LTG) of the KIG catalog. A comparison of the dark halo masses in the early- and late-type galaxies confirms our earlier results, as well as the findings of other authors—dark halos of isolated early-type galaxies turn out to be 2–3 times more massive than those of spiral galaxies of the same stellar mass.

The structure of the paper is as follows. Section 2 provides the results of the search for companions of the spiral galaxies in the KIG catalog. The main observed and calculated characteristics of LTG galaxies and their companions are given in Table 1 (Appendix). In Section 3 we describe a method for estimating the luminosity of spiral KIG galaxies in the $K$-band. Section 4 contains formulas for estimating the orbital masses of LTG galaxies. The concluding remarks are given in Section 5.

## 2. A SEARCH FOR COMPANIONS AROUND ISOLATED LTG GALAXIES

In our previous paper (Karachentseva et al. 2021) we presented a new morphological classification of isolated early-type galaxies (E, S0) from the KIG catalog, based on the data from up-to-date digital sky surveys. Moreover, a significant part of galaxies, 74 of 165, or 45%, passed into the category of spiral galaxies. We used a similar approach to clarify the classification of 885 late-type KIG galaxies. Five following galaxies among them: KIG 358, 533, 644, 782 and 952 have been reclassified as the E and S0-types.

Using the NED option, we searched for companions around 954 late-type KIG galaxies. As a result, it was found that 52% of them have small neighbors with a projection separation between the companion and the main galaxy of $R_p < 750$ kpc and the mutual radial velocity difference of less than 500 km s$^{-1}$. Note that the selection of galaxies into the KIG catalog by the condition of their isolation was applied to the galaxies in the Zwicky et al. (1961) catalog with an apparent photographic magnitude brighter than $15.7m$ and the declination north of $-2.5°$. To calculate the orbital masses, we limited ourselves to the value of $R_p < 330$ kpc, which approximately corresponds to the virial radius of a typical KIG galaxy, while maintaining the condition for the difference in radial velocities of $|\Delta V| < 500$ km s$^{-1}$. At that, we have excluded the nearby KIG galaxies with radial velocities below 1500 km s$^{-1}$ to avoid errors in the estimation of Hubble distances due to the effect of peculiar motions in the Local Supercluster.

Basic data on the KIG galaxies with companions are presented in Table 1 in the Appendix. The columns of the table list the following: (1)—the name of a galaxy in the LEDA or NED; (2)—the morphological type of a galaxy in the digital system we have determined from the PanSTARRS-1 survey; (3)—integral apparent $B$-magnitude $b_t$ from the LEDA; numbers with decimal places refer to our visual estimates; (4)—integral apparent $K$-band magnitude from the LEDA; (5)—integral $K$-band magnitude, determined from the $B$-magnitude and the morphological type (see Section 3); (6), (7)—radial velocity relative to the centroid of the Local Group and its error in km s$^{-1}$, adopted from the NED, with the addition for the companion of the KIG 853 from Melnyk et al. (2009); (8)—the difference between the radial velocities of the companion and the KIG galaxy; (9)—the distance modulus (modbest) in magnitudes from the LEDA; (10)—projection separation $R_p$, in kpc, between the companion and the KIG galaxy under the assumption that the companion distance is similar to that to the KIG galaxy; (11)—the logarithm of the $K$-band luminosity of a KIG galaxy, expressed in units of solar luminosity; (12)—the logarithm of the estimated orbital mass of the KIG galaxy in solar mass units (see Section 4); (13)—the difference of the logarithms of the orbital mass and the $K$-band luminosity.

Figure 1 shows the histogram of the distribution of galaxies by the magnitude difference between the companion and the KIG galaxy from Table 1. The average apparent magnitude difference is $\Delta B = 2.94^m \pm 0.15^m$. The presence of such faint companions does not violate the accepted isolation condition. However, in a few cases (for example, KIG 237 and CGCG 263-017), the apparent magnitude of the distant neighbor turned out to be comparable to the apparent magnitude of the KIG galaxy. In general, the detected companions can be considered as test particles that are making Keplerian motions around a massive central body.

Figure 2 shows the distribution of spiral KIG galaxies and their companions by the difference modulus between the radial velocities and the absolute $B$-band magnitude of the main galaxy. The median values of these parameters are, respectively, 101 km s$^{-1}$ and $-20.6^m$, which is typical of groups around the nearby massive spirals like our Galaxy, M 31 and M 81. The radial velocity difference shows a weak tendency to decrease with a decreasing luminosity of the KIG galaxies with the regression line $|dV| = -5.86 M_B + 19.96$ and the correlation coefficient $R = -0.057$. The presence of this correlation reflects the fact that the greater the mass of the central galaxy, the higher the spatial velocity of the companions. However, this expected relationship is blurred due to the effect of projection of the companion velocity onto the line of sight.

Figure 3 demonstrates a relationship between the radial velocity difference modulus of the companion relative to the KIG galaxy and the projection separation between them, $R_p$. About 90% of the companions have a radial velocity difference not exceeding 250 km s$^{-1}$, which corresponds to the typical rotation amplitude of massive spirals. We used the values of $|dV|$ and $R_p$ in 141 companions around 111 spiral KIG galaxies to calculate the total (orbital) masses of the KIG galaxies in Section 4.

3. $K$-BAND LUMINOSITY OF SPIRAL KIG GALAXIES

An important dynamic characteristic of a galaxy is the ratio of its integral stellar mass $M^*$ to the total mass $M_T$, the main contribution to which is made by the mass of the dark halo. Integral luminosity of a galaxy in the $K_s$-band, $L_K$ is usually taken as an optimal estimate for $M^*$. According to Bell et al. (2003), $L_K$ luminosity directly corresponds to the stellar mass: $M^*/L_K = 1.0 M_\odot/L_\odot$. Later research favors a slightly lower proportionality factor, $0.6 M_\odot/L_\odot$ (Lelli et al. 2016).

For many galaxies distributed throughout the sky, the apparent $K_s$ magnitudes were measured in the 2MASS survey (Skrutskie et al. 2006). Lyon Extragalactic Database = LEDA (Makarov et al. 2014) contains the apparent $K_t$ magnitudes of galaxies. They are based on the 2MASS survey data, supplemented, where available, with the $K$-magnitude estimates from other sources. As it is known (Jarrett 2000), due to short exposures the 2MASS survey is not deep and loses the flux from peripheral regions of galaxies, especially those having a bluish color. An underestimation of the flux of late-type galaxies can be quite significant. For this reason, Jarrett et al. (2003) proposed to estimate the $K$-magnitudes of galaxies by their $B$-magnitudes and the average color index $\langle B - K \rangle = 4.60 - 0.25 \times T$, depending on the morphological type $T = 2, 3...9$ in the de Vaucouleurs scale. For the early-type galaxies $T < 2$, the value of $\langle B - K \rangle$ was accepted to be equal to 4.10.

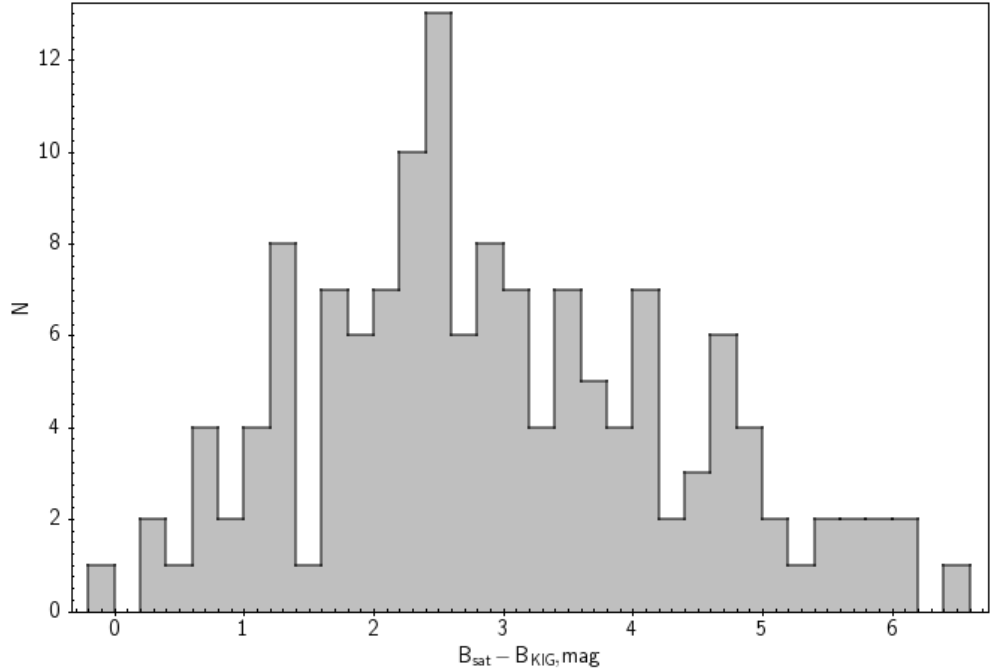

**Figure 1**: A histogram of the distribution of galaxies from Table 1 based on the magnitude difference between the companion and the KIG galaxy. The average difference in apparent magnitudes is $\Delta B = 2.94^m \pm 0.15^m$.

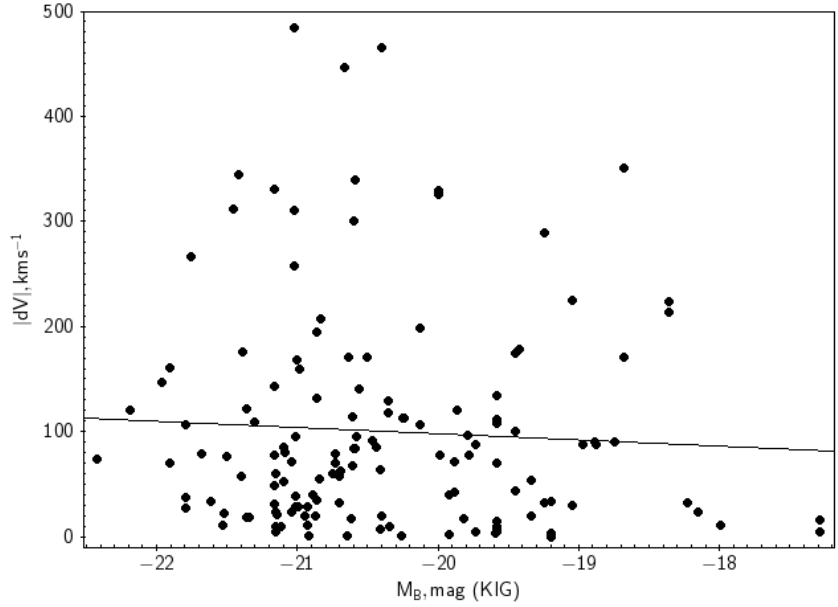

**Figure 2**: Modulus of the radial velocity difference 'companion–KIG galaxy' depending on the absolute magnitude of the KIG galaxy. The regression line (the straight line) is expressed as $|dV| = -5.86 M_B$ (KIG)$+19.96$; the correlation coefficient $R = -0.057$.

We denote the $K$-band magnitudes determined this way as $K_B$. The relationship between $K_B$ and $K_t$ from the LEDA for the KIG galaxies with the companions is presented in Fig. 4. Eighteen E and S0 galaxies from our previous paper (Karachentseva et al. 2021) were added to spiral galaxies, they are marked with crosses. As follows from these data, early-type galaxies are located near the diagonal line $K_t = K_B$ (dotted line), with the mean of $\langle K_t - K_B \rangle = 0.15^m \pm 0.06^m$ and the variance of $\sigma(K_t - K_B) = 0.24^m$. In the spiral KIG galaxies, $K_B$-magnitudes turn out to be systematically brighter than the $K_t$-magnitudes, following the regression line

$$K_t = (1.03 \pm 0.06) \times K_B + 0.44 \pm 0.62.$$

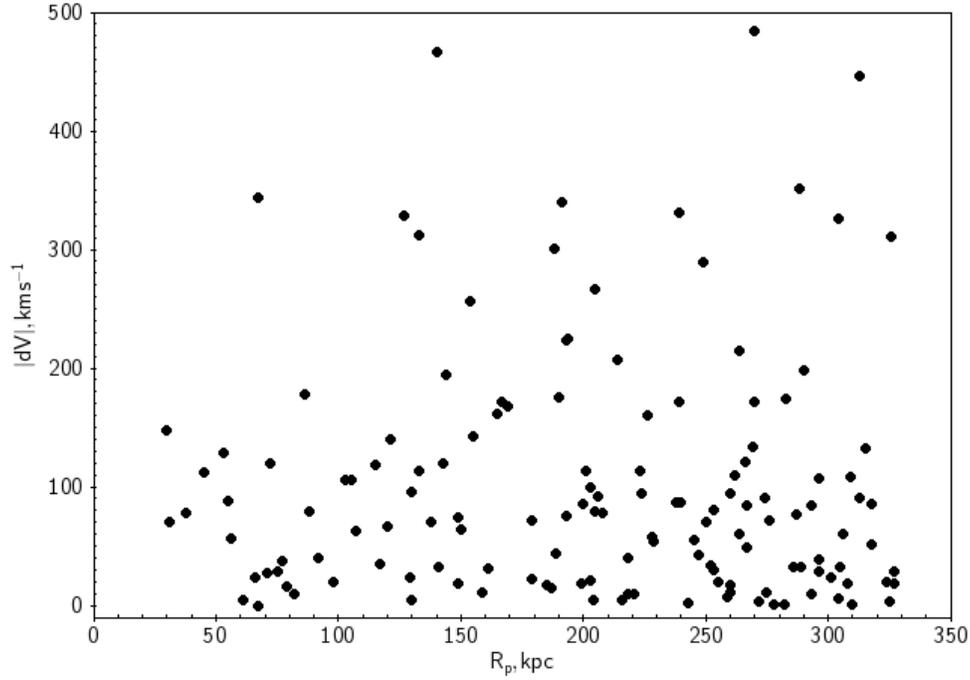

**Figure 3**: Relationship between the radial velocity difference modulus 'companion–KIG galaxy' and the projection separation $R_p$ between a companion and a KIG galaxy.

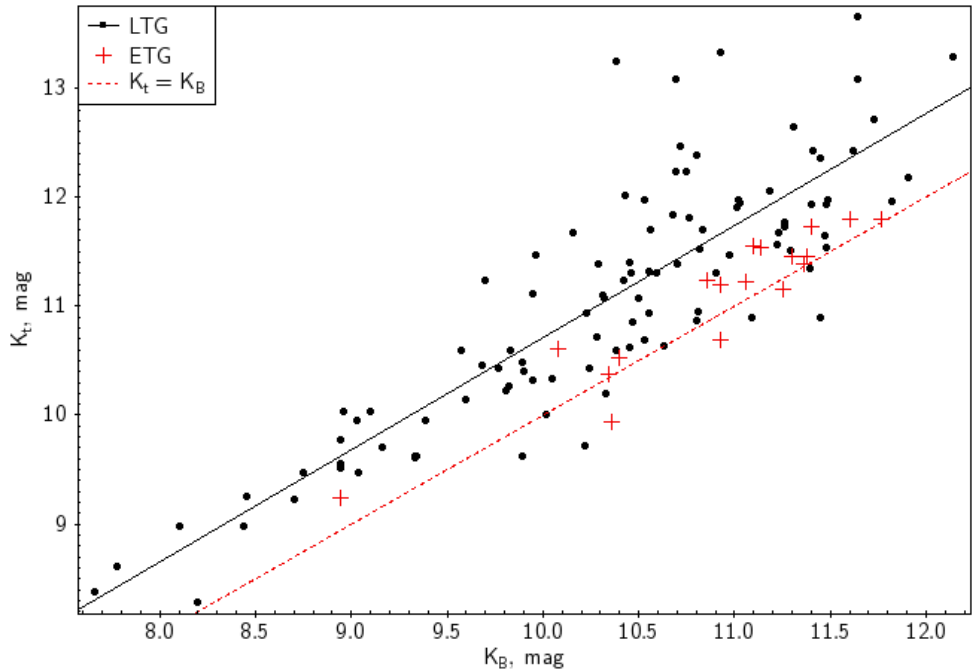

**Figure 4**: Relationship between the apparent magnitudes $K_B$ and $K_t$ for the KIG galaxies. The circles indicate late-type galaxies, the crosses—early-type galaxies (from the paper Karachentseva et al. (2021)). The dotted line corresponds to the equality of magnitudes $K_B = K_t$. The solid line describes the regression $K_t = (1.03 \pm 0.66)K_B + (0.44 \pm 0.62)$. The correlation coefficient is $R = 0.86$.

As we can see from the figure, the $K_t - K_B$ difference and its variance show a tendency to increase from the bright to faint galaxies. The average $\langle K_t - K_B \rangle$ difference for late-type galaxies is $+0.71^m \pm 0.07^m$.

We have taken the data on the $B$-magnitudes from the LEDA, as corrected by the LEDA for the Galactic and internal extinction. It should be noted here that the value of the internal extinction in terms of the LEDA algorithm seems to us somewhat overestimated, especially for the low luminosity galaxies (Melnyk et al. 2017). This circumstance makes it possible to explain a part of the observed systematic difference between $K_B$ and $K_t$.

## 4. FINDING ORBITAL MASSES FOR LATE-TYPE KIG GALAXIES

In the presence of an ensemble of test particles around a compact attractor, the estimate of its mass, $M$, by the orbital motions of the companions can be expressed as

$$M_{\rm orb} = \langle \eta_e \rangle^{-1} G^{-1} \langle \Delta V^2 \times R_p \rangle, \quad (1)$$

where $G$ is the constant of gravitation, $\Delta V$ is the difference in radial velocities of the companion and the central object, $R_p$—their linear projection separation, and $\langle \eta_e \rangle$ is the average value of the dimensionless projection factor, which depends on the eccentricity of the orbit of the companions. In the case of elliptical Keplerian motions with the orbit eccentricity $e$, the projection factor has the form

$$\begin{aligned}\eta_e(i, \Omega, \omega) &= \sin^2 i [1 - \sin^2 i \sin^2 \Omega]^{1/2} \\ &\times [e \cos(\Omega - \omega) + \cos \Omega]^2 (1 + e \cos \omega)^{-1}.\end{aligned} \quad (2)$$

Here $\omega$ means the angle between the major axis of the orbit and the line of nodes, $\Omega$ is the angle between the line connecting the objects and the line of nodes, and $i$ is the angle of inclination of the orbital plane to the plane of the sky. With a chaotic orientation of the companion orbits, the distribution density of the three angles is

$$\begin{aligned}P_e(i, \Omega, \omega) &= (1 - e^2)^{1/2} \\ &\times \sin i (1 + e \cos \omega)^2 / 4\pi^2,\end{aligned} \quad (3)$$

where $[0 \leq i \leq \pi/2, 0 \leq \Omega \leq 2\pi, 0 \leq \omega \leq 2\pi]$. Then, according to Karachentsev (1987),

$$\langle \eta_e \rangle = (3\pi/32)(1 - 2e^2/3), \quad (4)$$

$$\langle \eta_e^2 \rangle = (6/35)(1 - 5e^2/6). \quad (5)$$

A simulation of the orbits of companions of a massive galaxy by Barber et al. (2014), shows that for the ensemble of orbits we can take the value of $\langle e^2 \rangle = 1/2$. Then the estimate of the orbital mass of the central galaxy has the form

$$\begin{aligned}M_{\rm orb} &= (16/\pi) G^{-1} \langle \Delta V^2 R_p \rangle \\ &= 1.18 \times 10^6 \langle \Delta V^2 R_p \rangle,\end{aligned} \quad (6)$$

where $\Delta V$ is expressed in $\rm km\,s^{-1}$, $R_p$—in kpc, and mass—in solar mass units. Individual values of $M_{\rm orb}$, obtained from the difference of radial velocities and projection separations of 141 companions, are presented in the Appendix (in Table 1).

We determined the integral luminosity of KIG galaxies in the $K$-band from the apparent $K_B$-magnitudes

$$\log(L_K/L_\odot) = 0.4(3.28 + \mu - K_B), \quad (7)$$

where $\mu$ is the galaxy distance modulus ('modbest') given in the LEDA, and 3.28 is the absolute magnitude of the Sun in the $K$-band. For the galaxies beyond the boundaries of the Local Supercluster, the distance modulus approximately corresponds to the Hubble parameter $H_0 = 70 \rm \ km\,s^{-1} Mpc^{-1}$.

The distribution of KIG galaxies by the orbital mass and the integral $K$-band luminosity is shown in Fig. 5. The empty circles at the top of the figure represent objects with an abnormally

7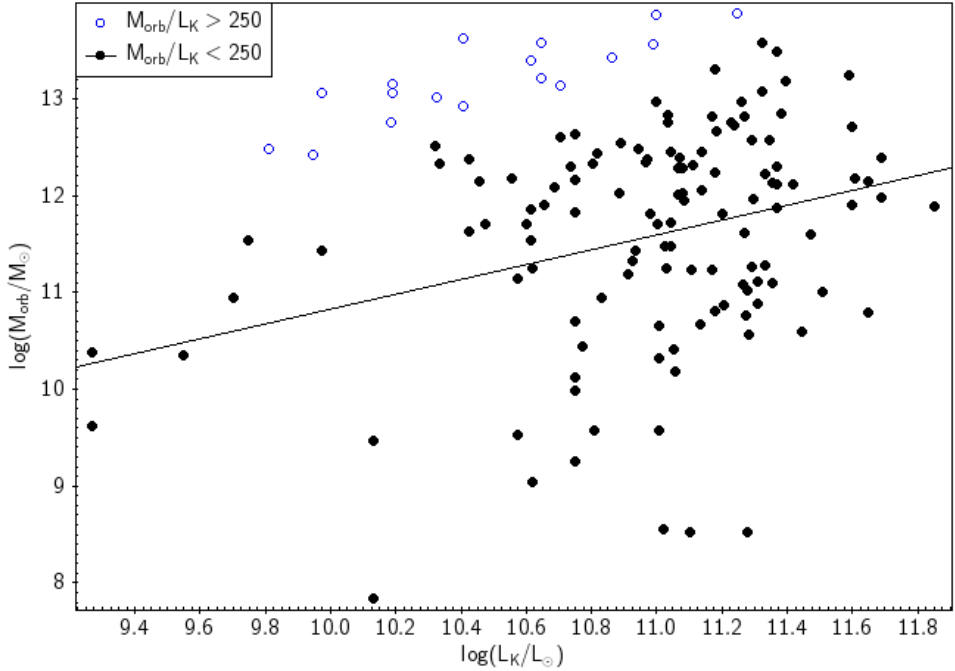

**Figure 5**: Relationship in the logarithmic scale between $M_{\rm orb}$ and $L_K$. Unfilled circles refer to the KIG galaxies with the ratio $M_{\rm orb}/L_K > 250$ in solar units, which are presumably included in the diffuse filaments and groups. For the rest of the galaxies, there happen to be a positive correlation $\log(M_{\rm orb}/M_\odot) = (0.77 \pm 0.21)\log(L_K/L_\odot) + 3.15 \pm 2.28$ with a correlation coefficient $R = 0.32$.

large $M_{\rm orb}/L_K > 250 M_\odot/L_\odot$ ratio (see below). The figure illustrates the presence of a positive correlation between $M_{\rm orb}$ and $L_K$, which is described by the linear regression $\log(M_{\rm orb}/M_\odot) = (0.77 \pm 0.21)\log(L_K/L_\odot) + 3.15 \pm 2.28$ (excluding abnormal cases). The relationship between orbital and stellar masses is subject to a large dispersion due to the geometric projection effect. The median values of both parameters for our sample are approximately $1.0 \times 10^{12} M_\odot$ and $1.0 \times 10^{11} L_\odot$, which is comparable to the mass and luminosity of the Milky Way and Andromeda (M 31).

The histogram in Fig. 6 (black line) represents the distribution of KIG galaxies by the magnitude of the $M_{\rm orb}/L_K$ ratio in the logarithmic scale. The distribution extends for 6 orders of magnitude and looks asymmetrical with a maximum at about $15 M_\odot/L_\odot$. To estimate the role of the projection effect, affecting the shape of the distribution $N[\log(M_{\rm orb}/L_K)]$, we used the expressions for the projection factor $\eta$ and its probability density $p(\eta)$, indicated above. The red histogram in the figure shows the expected distribution of $M_{\rm orb}/L_K$ estimates for a fixed eccentricity of companion orbits $e = 1/\sqrt{2} \simeq 0.7$ and a fixed $M_T/L_K = 25 M_\odot/L_\odot$ ratio. It was built using the method of random testing. The modeled distribution has an asymmetry similar to that of the observed distribution, but with a sharper peak at the maximum. In this case, the maximum simulated value is $M_{\rm orb}/L_K = 128 M_\odot/L_\odot$. Assuming the presence of variance in the $M_T/L_K$ values for KIG galaxies, $\sigma[\log(M_T/L_K)] \simeq 0.3$, we can achieve a better fit between the simulated and observed distributions in the peak region. However, in the area of $M_{\rm orb}/L_K > 250 M_\odot/L_\odot$ values there remains a significant misalignment. The obvious reason for it is an admixture of fictitious systems formed by the KIG galaxies and dwarfs belonging to diffuse elements, e.g. the filaments of the Large-Scale Structure of the Universe. The KIG 32+KIG 34 and KIG 56+KIG 60 pairs can serve as examples to that, where the KIG galaxies are associated with each other by the coordinates and radial velocities. The relative number of such cases is small, $17/141 \simeq 12\%$, but their contribution to the average mass estimate turns out to be essential. Excluding the systems with $M_\mathrm{orb}/L_K > 250 M_\odot/L_\odot$ as non physical pairs, we obtain for the spiral KIG galaxies the average value of $\langle M_\mathrm{orb}/L_K \rangle = (20.9 \pm 3.1) M_\odot/L_\odot$.

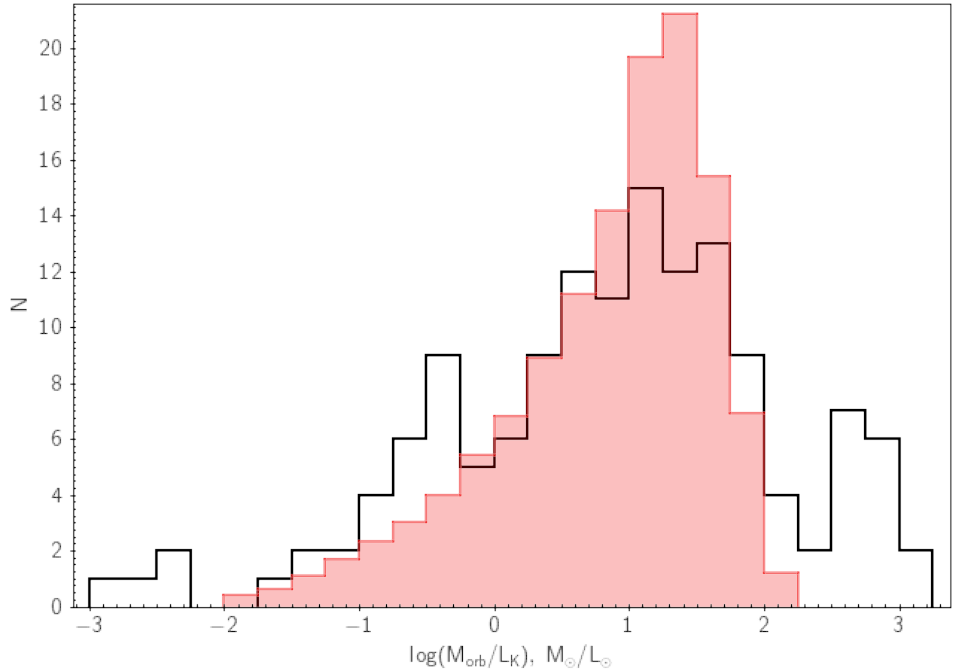

**Figure 6**: Distribution of KIG galaxies according to the estimates of the $M_\mathrm{orb}/L_K$ ratio in the logarithmic scale (the black histogram). The red histogram shows the expected distribution of $M_\mathrm{orb}/L_K$ at the fixed $M_T/L_K = 25 M_\odot/L_\odot$ ratio and the eccentricity of the orbits of companions $e = 0.7$.

As we can see from Fig. 1, there are 10 companions in our sample, their luminosity is comparable to the luminosity of the KIG galaxy itself ($B_\mathrm{sat} - B_\mathrm{KIG} < 1.0^m$). Their typical projection separation is $R_p \simeq 240$ kpc. The exclusion of these cases insignificantly changes the average $\langle M_\mathrm{orb}/L_K \rangle$ estimate, from $20.9 \pm 3.1$ to $20.4 \pm 3.0$ in the units of mass and luminosity of the Sun.

It should be noted that the character of the galaxy distribution by the $M_\mathrm{orb}/L_K$ parameter is affected by some other factors. The motions of dwarf companions in the extended halos of central massive galaxies differ from the Keplerian ones. On the average, taking into account the extension of dark halos slightly increases the $\langle M_\mathrm{orb}/L_K \rangle$ estimate. On the other hand, radial velocities of galaxies are measured with a noticeable error. Hence, the mean square error of the velocity difference for the KIG galaxies and their companions amounts to $\sigma(\Delta V) = 26$ km s$^{-1}$. Due to the quadratic dependence of finding $M_\mathrm{orb}$ on $\Delta V$, this overestimates the average value of the mass estimate by about 12%, and also affects the shape of the distribution tail at $M_\mathrm{orb}/L_K < 1$.

We estimated the stellar mass of KIG galaxies from their $K$-band luminosity, recalculated from the $B$-band luminosity taking into account the morphological type according to the method from Jarrett et al. (2003). Figure 4 shows that in the considered sample, there are 26 KIG galaxies with the $K_t$-magnitudes from the LEDA much fainter than the $K_B$ values we used, namely, $K_t - K_B > 1.0^m$. Four galaxies stand out among them: KIG 476, KIG 495, KIG 502 and KIG 949 with $K_t - K_B > 2.0^m$. Analysis of these cases shows that these galaxies have an extended peripheral structure of low surface brightness, not registered by the 2MASS survey. The average difference of $B$ and $K$-magnitudes for them from the LEDA is $\langle b_t - k_t \rangle = 1.58^m \pm 0.33^m$, which is not typical for spiral galaxies. Note, however, that exclusion of 26 galaxies with a large difference in the $K$- band magnitude estimates has little effect on the $M_{\rm orb}/L_K = 20.9 \pm 3.1$ estimate, decreasing the value to $18.9 \pm 2.8 M_\odot/L_\odot$.

**Table 2**: The average ratio of the orbital mass to the $K$-band luminosity for isolated galaxies of different types

| Type | $T < 0$ | 0–2 | 3, 4 | 5–8 | All |
|---|---|---|---|---|---|
| N(sat) | 31 | 42 | 38 | 39 | 150 |
| $\langle M_{\rm orb}/L_K \rangle$ | 59.8 $\pm$22.9 | 21.3 $\pm$6.1 | 16.3 $\pm$3.6 | 24.9 $\pm$5.4 | 28.9 $\pm$5.1 |

As noted by many authors, the $M_T/M^*$ or $M_T/L_K$ ratio of the galaxies with predominant bulges is on the average distinctively larger than that in disk-shaped galaxies. According to the data, by Karachentseva et al. (2011), Mandelbaum et al. (2016), More et al. (2011) this difference reaches a factor of about 2–4, depending on the way in which the $M_T/M^*$ is estimated. We have expanded our sample of 124 companions around the spiral KIG galaxies with the estimates of $M_{\rm orb}/L_K < 250 M_\odot/L_\odot$ with other 26 companions around the E and S0 KIG-galaxies from Karachentseva et al. (2021). In doing so, we took into account the $\langle K_B - K_t \rangle = 0.15^m$ correction for the transition to the system of $K_B$-magnitudes we have adopted. The KIG galaxies were combined into four subgroups according to their morphological type: E, S0 ($T < 0$), S0a, Sa, Sab ($T = 0$–2), Sb, Sbc ($T = 3, 4$) and Sc–Sm ($T = 5$–8). The number of companions in each subgroup, the average $M_{\rm orb}/L_K$ values and mean errors are presented in Table 2. As we can see from Fig. 7, reproducing the data of this table, spiral galaxies of different types have similar values of the halo mass-to-stellar mass ratio within the statistical error. However, for the E and S0 galaxies the average $\langle M_{\rm orb}/L_K \rangle = 59.8 \pm 22.9 M_\odot/L_\odot$ ratio turns out to be about 2 times higher than that of the spiral galaxies.

We noted a similar result for the isolated galaxies of the 2MIG catalog (Karachentseva et al. 2010), compiled using the 2MASS survey (Skrutskie et al. 2006). Based on the radial velocity and projection separation data for 154 small companions around the spiral 2MIG galaxies, Karachentseva et al. (2011) found the median ratio $M_{\rm orb}/L_K = 17 M_\odot/L_\odot$, while based on the motions of 60 companions around the E and S0 galaxies, the median ratio turned out to be 63 $M_\odot/L_\odot$.

## 5. CONCLUDING REMARKS

The initial morphological classification in the Catalog of Isolated Galaxies (Karachentseva 1973) was carried out based on the reproductions of the photographic Palomar Observatory Sky Survey POSS-I. We have revised the old classification, based on the modern digital multicolor PanSTARRS-I (Chambers et al. 2016) and SDSS (Ahn et al. 2014) sky surveys, also using the ultraviolet and infrared range data. As a result, among 1050 KIG galaxies, the number of early-type objects (E and S0) has decreased from 165 to 91 (Karachentseva et al. 2021). At the same time, the updated classification of 959 spiral KIG galaxies discussed here has reclassified only 5 galaxies into the E and S0 category.

As a result of our search for the companions with measured radial velocities, we have discovered 31 companion around 23 early-type KIG galaxies and 136 companions around 106 late-type KIG galaxies within the projection separation of 330 kpc, which roughly corresponds to the virial radius of a typical KIG galaxy. The restriction on the 'KIG–companion' radial velocity difference amounted to 500 km s$^{-1}$. Therefore, about 76% of isolated ETG galaxies proved to be without any visible companions, while among the LTG galaxies this value was 89%. This indicates a good efficiency of the isolation criterion, used in the KIG catalog. The discovered companions of the early-type KIG galaxies turned out to be, on the average, weaker than the 'host' galaxies by $\Delta B = 2.53^m \pm 0.15^m$, and for the companions of spiral KIG galaxies, the average apparent magnitude difference amounted to $\Delta B = 2.94^m \pm 0.15^m$. Having such faint companions practically does not violate the isolation condition adopted in the KIG catalog.

Considering these small companions as test particles, moving around the KIG galaxies on the Keplerian orbits with the eccentricity of $e \simeq 0.7$, we have estimated the total masses of KIG galaxies. For the isolated galaxies of all morphological types we have obtained the average ratio $\langle M_{\rm orb}/L_K \rangle = (28.9 \pm 5.1) M_\odot/L_\odot$, which matches the average $(31 \pm 6) M_\odot/L_\odot$ ratio for 25 bright galaxies of the Local Volume, similar to the Milky Way and M 31 (Karachentsev and Kudrya 2014). Moreover, the $\langle M_{\rm orb}/L_K \rangle$ ratio for the E and S0 KIG galaxies turns out to be about twice as high $(59.8 \pm 22.9) M_\odot/L_\odot$ than that for the spiral galaxies of the same stellar mass, $(20.9 \pm 3.1) M_\odot/L_\odot$, which is also consistent with the data for massive galaxies of the Local Volume. The systematic difference in the stellar mass-to-halo mass ratio in the early and late-type galaxies obviously indicates a significant difference in the character of dynamic evolution of the disk-shaped and bulge-shaped galaxies.

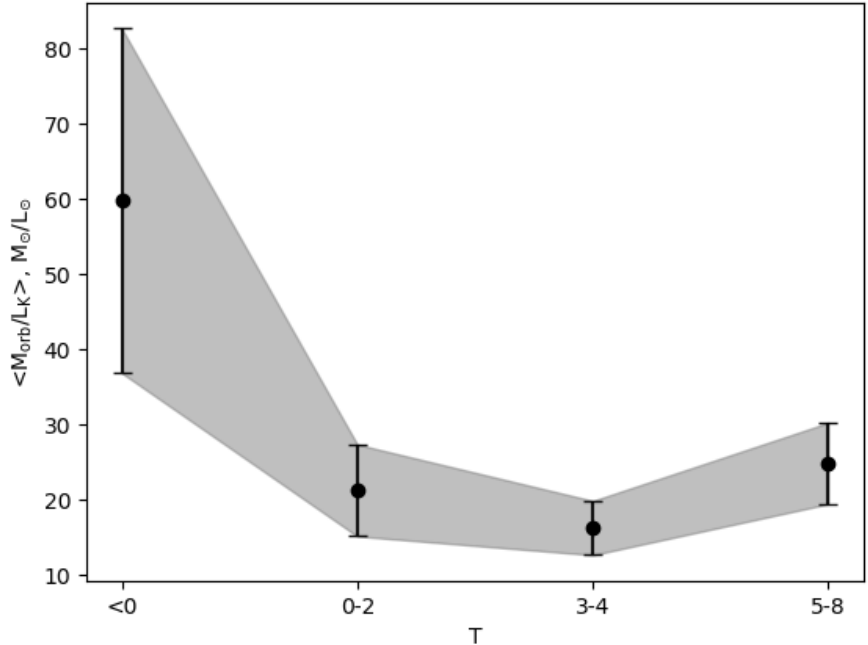

**Figure 7**: Dependence of the $M_{\rm orb}/L_K$ ratio on the morphological type of isolated galaxy. The KIG galaxies were combined into four subgroups according to their morphological type: E, S0 ($T < 0$); S0a, Sa, Sab ($T = 0, 1, 2$); Sb, Sbc ($T = 3, 4$) and Sc–Sm ($T = 5, 6, 7, 8$). Early-type galaxies are characterized by a dark halo, which is approximately 2-3 times more massive than the halo of late-type galaxies having the same stellar mass.

As follows from the data in Fig. 6, the observed distribution of KIG galaxies by $M_{\rm orb}/L_K$ has some excess over the expected distribution near $M_{\rm orb}/L_K \simeq 0.5 M_\odot/L_\odot$. This may indicate the presence of a not numerous subsystem of spiral galaxies (about 10%), lacking massive dark halos. Analyzing the kinematics of companions around 25 nearby spiral galaxies with the Milky Way-type luminosity, Karachentsev and Kashibadze (2021) noted the presence among them of 5 galaxies with an average ratio $\langle M_{\rm orb}/L_K \rangle = (5.8 \pm 1.1) M_\odot/L_\odot$. All these galaxies (NGC 253, NGC 2683, NGC 2903, NGC 3521 and NGC 5055) reveal decreasing rotation curves observed at the periphery, which also indicates the low mass of their halo. The existence of spiral galaxies without dark halos could be a serious challenge for the standard cosmological model.

**Acknowledgements.** The authors thank the anonymous referee for helpful advice. We made use of the PanSTARRS-I, SDSS, 2MASS sky surveys as well as the HyperLEDA (http://leda.univ-lyon1.fr) and NED (http://ned.ipac.caltech.edu) databases. This study was supported in part by the Russian Science Foundation (grant no. 19-12-00145).

APPENDIX

Table 1: Properties of the LTG KIG galaxies and their companions

| Galaxy name | $T$ | $b_t$ | $K_t$ | $K_B$ | $V_{LG}$ | $\sigma V_{LG}$ | $\Delta V_{LG}$ | modbest | $R_p$ | $\log(L_K)$ | $\log(M_{\rm orb})$ | $\log M_{\rm orb} - \log L_K$ |
|---|---|---|---|---|---|---|---|---|---|---|---|---|
| | | mag | mag | mag | km s$^{-1}$ | km s$^{-1}$ | km s$^{-1}$ | mag | kpc | L$_\odot$ | M$_\odot$ | |
| (1) | (2) | (3) | (4) | (5) | (6) | (7) | (8) | (9) | (10) | (11) | (12) | (13) |
| **KIG 16** | 0 | 15.55 | 11.68 | 11.23 | 5634 | 2 | | 34.49 | | | | |
| SDSS J002220.83+010734.7 | 9 | 19.72 | | | 5345 | 1 | -289 | | 249 | 10.616 | 13.390 | 2.774 |
| PGC 1181260 | 10 | 18.18 | 15.58 | | 5666 | 6 | 32 | | 286 | 10.616 | 11.540 | 0.924 |
| **KIG 19** | 0 | 15.31 | 11.30 | 10.90 | 5601 | 15 | | 34.49 | | | | |
| WISEA J002417.59+141619.6 | -2 | 16.67 | 12.68 | | 5713 | 52 | 112 | | 45 | 10.748 | 11.823 | 1.075 |
| WISEA J002425.18+141621.1 | 0 | 18.08 | | | 5596 | 45 | -5 | | 61 | 10.748 | 9.255 | -1.493 |
| SDSS J002431.06+141538.1 | 7 | 17.8 | | | 5611 | 6 | 10 | | 82 | 10.748 | 9.987 | -0.761 |
| 2MASX J00234382+1418246 | 5 | 16.06 | 13.70 | | 5586 | 6 | -15 | | 187 | 10.748 | 10.699 | -0.049 |
| SDSS J002443.46+140419.4 | 10 | 17.99 | | | 5531 | 9 | -70 | | 250 | 10.748 | 12.158 | 1.410 |
| AGC 748790 | 10 | 19.5 | | | 5607 | 4 | 6 | | 304 | 10.748 | 10.114 | -0.634 |
| WISEA J002500.25+140407.0 | 9 | 17.73 | | | 5493 | 2 | -108 | | 309 | 10.748 | 12.628 | 1.880 |
| **KIG 32** | 2 | 14.28 | 11.47 | 9.96 | 5526 | 2 | | 34.46 | | | | |
| AGC 102267 | 8 | 18.30 | 13.82 | | 5434 | 4 | -92 | | 206 | 11.112 | 12.314 | 1.202 |
| **KIG 34** | 8 | 15.68 | 13.28 | 12.14 | 5188 | 3 | | 34.32 | | | | |
| AGC 101825 | 10 | 19.2 | | | 5322 | 3 | 134 | | 269 | 10.184 | 12.756 | 2.872 |
| **KIG 56** | 3 | 14.06 | 10.43 | 9.77 | 5272 | 1 | | 34.33 | | | | |
| WISEA J013508.78+004303.7 | 9 | 17.85 | 15.04 | | 5332 | 2 | 60 | | 264 | 11.136 | 12.049 | 0.913 |
| **KIG 60** | 4 | 15.41 | 12.06 | 11.18 | 5256 | 3 | | 34.33 | | | | |
| PGC 1147952 | 7 | 18.00 | 13.80 | | 5256 | 3 | 3 | | 325 | 10.572 | 9.531 | -1.041 |
| **KIG 72** | 3 | 14.43 | 10.32 | 9.95 | 5284 | 8 | | 34.35 | | | | |
| IC 1749 | 9 | 14.69 | 11.07 | | 5200 | 8 | -84 | | 293 | 11.072 | 12.387 | 1.315 |
| **KIG 74** | 4 | 14.94 | 11.91 | 11.01 | 10400 | 7 | | 35.88 | | | | |
| IC 1752 | 3 | 16.29 | 12.44 | | 10744 | 3 | 344 | | 67 | 11.260 | 12.971 | 1.711 |
| **KIG 81** | 1 | 13.91 | 10.03 | 8.96 | 3634 | 41 | | 33.51 | | | | |
| WISEA J020001.90+123218.2 | 8 | 18.30 | | | 3654 | 2 | 20 | | 98 | 11.132 | 10.663 | -0.469 |

**Table 1**: (Contd.)

| Galaxy name | T | $b_t$ | $K_t$ | $K_B$ | $V_{LG}$ | $\sigma V_{LG}$ | $\Delta V_{LG}$ | modbest | $R_p$ | $\log(L_K)$ | $\log(M_{\rm orb})$ | $\log M_{\rm orb} - \log L_K$ |
|---|---|---|---|---|---|---|---|---|---|---|---|---|
| | | mag | mag | mag | km s$^{-1}$ | km s$^{-1}$ | km s$^{-1}$ | mag | kpc | L$_\odot$ | M$_\odot$ | |
| (1) | (2) | (3) | (4) | (5) | (6) | (7) | (8) | (9) | (10) | (11) | (12) | (13) |
| **KIG 86** | 3 | 13.88 | 10.14 | 9.6 | 5466 | 3 | | 34.44 | | | | |
| AGC 122960 | 9 | 16.56 | | | 4982 | 4 | -484 | | 270 | 11.248 | 13.873 | 2.625 |
| **KIG 96** | 5 | 11.62 | 8.61 | 7.78 | 1663 | 2 | | 31.05 | | | | |
| AGC 121171 | 9 | 16.38 | 13.70 | | 1703 | 3 | 40 | | 92 | 10.620 | 11.240 | 0.620 |
| KKH 10 | 10 | 16.5 | | | 1661 | 2 | -2 | | 243 | 10.620 | 9.041 | -1.579 |
| **KIG 120** | 2 | 14.63 | 10.00 | 10.02 | 6836 | 5 | | 34.95 | | | | |
| AGC122903 | 8 | 18.5 | | | 6847 | 4 | 11 | | 260 | 11.284 | 10.568 | -0.716 |
| **KIG 150** | 5 | 15.50 | 12.71 | 11.73 | 9174 | 5 | | 35.61 | | | | |
| PGC 146300 | 5 | 16.28 | 12.67 | | 8834 | 39 | -340 | | 191 | 10.864 | 13.416 | 2.552 |
| **KIG 159** | 5 | 15.45 | 9.63 | 9.89 | 4310 | 8 | | 34.20 | | | | |
| MCG +13-5-3 | -2 | 16.30 | 11.64 | | 4470 | 25 | 160 | | 226 | 11.036 | 12.834 | 1.798 |
| **KIG 181** | 5 | 14.04 | 10.94 | 10.23 | 4263 | 4 | | 33.99 | | | | |
| VV 640 | 9 | 15.59 | | | 4348 | 60 | 85 | | 318 | 10.816 | 12.433 | 1.617 |
| **KIG 198** | 5 | 15.64 | 11.35 | 11.39 | 9681 | 9 | | 35.77 | | | | |
| KUG 0734+464 | 9 | 17.32 | 14.18 | | 9601 | 2 | -80 | | 253 | 11.064 | 12.281 | 1.217 |
| **KIG 199** | 5 | 14.86 | 11.30 | 10.46 | 3896 | 2 | | 33.80 | | | | |
| WISEA J073913.79+374037.9 | 8 | 18.20 | | | 3567 | 1 | -329 | | 127 | 10.648 | 13.210 | 2.562 |
| WISEA J073951.88+374915.6 | 7 | 18.2 | | | 3570 | 2 | -326 | | 304 | 10.648 | 13.581 | 2.933 |
| **KIG 211** | 3 | 15.61 | 10.87 | 10.8 | 8345 | 3 | | 35.44 | | | | |
| WISEA J075330.46+323046.3 | 5 | 18.46 | | | 8380 | 4 | 35 | | 117 | 11.168 | 11.228 | 0.06 |
| WISEA J075312.13+323647.1 | 7 | 17.53 | | | 8150 | 2 | -195 | | 144 | 11.168 | 12.810 | 1.642 |
| **KIG 213** | 0 | 14.80 | 10.62 | 10.45 | 5952 | 3 | | 34.72 | | | | |
| WISEA J075528.53+392512.7 | 9 | 17.18 | 14.17 | | 5951 | 2 | -1 | | 310 | 11.020 | 8.556 | -2.464 |
| **KIG 232** | 5 | 13.06 | 9.61 | 9.33 | 5215 | 2 | | 34.44 | | | | |
| SDSS J081025.18+340015.5 | 9 | 16.71 | | | 5184 | 2 | -37 | | 77 | 11.356 | 11.093 | -0.263 |
| SDSS J081021.15+340158.7 | 8 | 18.05 | | | 5321 | 2 | 106 | | 103 | 11.356 | 12.134 | 0.778 |
| **KIG 237** | 5 | 15.23 | 9.72 | 10.22 | 5525 | 3 | | 34.55 | | | | |

**Table 1**: (Contd.)

| Galaxy name | T | $b_t$ | $K_t$ | $K_B$ | $V_{LG}$ | $\sigma V_{LG}$ | $\Delta V_{LG}$ | modbest | $R_p$ | $\log(L_K)$ | $\log(M_{\text{orb}})$ | $\log M_{\text{orb}} - \log L_K$ |
|---|---|---|---|---|---|---|---|---|---|---|---|---|
| | | mag | mag | mag | km s$^{-1}$ | km s$^{-1}$ | km s$^{-1}$ | mag | kpc | L$_\odot$ | M$_\odot$ | |
| (1) | (2) | (3) | (4) | (5) | (6) | (7) | (8) | (9) | (10) | (11) | (12) | (13) |
| WISEA J081507.61+523439.8 | -5 | 18.16 | | | 5620 | 12 | 95 | | 260 | 11.044 | 12.442 | 1.398 |
| PGC 2425695 | -2 | 17.30 | | | 5554 | 12 | 29 | | 296 | 11.044 | 11.468 | 0.424 |
| CGCG 263-017 | 3 | 15.19 | 10.81 | | 5564 | 3 | 39 | | 296 | 11.044 | 11.725 | 0.681 |
| **KIG 250** | 8 | 13.63 | 11.38 | 10.29 | 2157 | 5 | | 31.27 | | | | |
| MCG +08-16-005 | 9 | 16.08 | 13.21 | | 2181 | 1 | 24 | | 129 | 9.704 | 10.944 | 1.240 |
| **KIG 252** | 5 | 15.50 | 12.64 | 11.31 | 9369 | 7 | | 35.70 | | | | |
| WISEA J082622.45+112203.6 | 7 | 18.2 | | | 9421 | 9 | 52 | | 318 | 11.068 | 12.004 | 0.936 |
| **KIG 270** | 2 | 15.40 | 11.98 | 10.53 | 8849 | 15 | | 35.58 | | | | |
| WISEA J083821.75+124354.1 | 9 | 18.34 | | | 8920 | 2 | 71 | | 276 | 11.332 | 12.215 | 0.883 |
| WISEA J083825.49+124429.7 | -5 | 16.65 | 12.57 | | 8872 | 4 | 23 | | 301 | 11.332 | 11.274 | -0.058 |
| **KIG 274** | 1 | 15.78 | 11.54 | 11.48 | 7462 | 7 | | 35.22 | | | | |
| WISEA J084017.68+231058.0 | 7 | 18.29 | | | 7468 | 2 | 4 | | 204 | 10.808 | 9.580 | -1.228 |
| **KIG 278** | 3 | 14.45 | 11.12 | 9.95 | 7661 | 6 | | 35.28 | | | | |
| SDSSJ084131.79+325932.0 | 9 | 17.89 | | | 7672 | 4 | 11 | | 275 | 11.444 | 10.591 | -0.853 |
| **KIG 293** | 7 | 14.98 | | 10.58 | 1719 | 13 | | 32.16 | | | | |
| PGC1746243 | 9 | 17.46 | | | 1809 | 0 | 90 | | 274 | 9.944 | 12.418 | 2.474 |
| **KIG 296** | 4 | 14.60 | 10.80 | 9.86 | 4244 | 3 | | 34.01 | | | | |
| PGC 1917858 | 0 | 15.20 | 11.27 | | 4149 | 2 | -95 | | 224 | 10.972 | 12.377 | 1.405 |
| **KIG 319** | 3 | 14.16 | 9.95 | 9.39 | 4802 | 4 | | 34.34 | | | | |
| WISEA J091529.31+114625.3 | 9 | 17.42 | | | 4945 | 15 | 143 | | 155 | 11.292 | 12.573 | 1.281 |
| WISEA J091508.03+114550.9 | 1 | 17.06 | | | 4771 | 3 | -31 | | 161 | 11.292 | 11.262 | -0.03 |
| **KIG 323** | 6 | 15.15 | 11.40 | 10.45 | 7910 | 4 | | 35.35 | | | | |
| SDSS J092049.21+241619.0 | 9 | 18.48 | | | 7892 | 2 | -18 | | 149 | 11.272 | 10.756 | -0.516 |
| **KIG 338** | 0 | 14.53 | 10.43 | 10.24 | 3294 | 3 | | 33.49 | | | | |
| AGC 193045 | 9 | 16.97 | | | 3328 | 2 | 34 | | 252 | 10.612 | 11.536 | 0.924 |
| **KIG 339** | 3 | 14.52 | 10.85 | 10.47 | 7610 | 2 | | 35.26 | | | | |
| WISEA J092948.09+554642.4 | 8 | 18.20 | | | 7778 | 2 | 168 | | 169 | 11.228 | 12.750 | 1.522 |

**Table 1**: (Contd.)

| Galaxy name | T | $b_t$ | $K_t$ | $K_B$ | $V_{LG}$ | $\sigma V_{LG}$ | $\Delta V_{LG}$ | modbest | $R_p$ | $\log(L_K)$ | $\log(M_{\rm orb})$ | $\log M_{\rm orb} - \log L_K$ |
|---|---|---|---|---|---|---|---|---|---|---|---|---|
| | | mag | mag | mag | km s$^{-1}$ | km s$^{-1}$ | km s$^{-1}$ | mag | kpc | L$_\odot$ | M$_\odot$ | |
| (1) | (2) | (3) | (4) | (5) | (6) | (7) | (8) | (9) | (10) | (11) | (12) | (13) |
| **KIG 348** | 4 | 16.18 | 13.48 | 11.84 | 11016 | 2 | | 36.07 | | | | |
| WISEA J093317.15+053546.6 | 9 | 17.99 | | | 10953 | 1 | -63 | | 107 | 11.004 | 11.700 | 0.696 |
| **KIG 355** | 7 | 14.43 | 10.95 | 10.81 | 4337 | 2 | | 34.07 | | | | |
| WISEA J093617.01+373751.3 | 9 | 17.83 | | | 4401 | 1 | 64 | | 150 | 10.616 | 11.860 | 1.244 |
| **KIG 358** | -2 | 13.30 | 9.47 | 9.04 | 3658 | 44 | | 33.71 | | | | |
| AGC 193900 | 7 | 18.5 | | | 3675 | 4 | 17 | | 185 | 11.180 | 10.799 | -0.381 |
| **KIG 362** | 3 | 15.46 | 11.52 | 10.82 | 9489 | 7 | | 35.73 | | | | |
| WISEA J094259.81+630630.5 | 9 | 17.47 | | | 9468 | 2 | -21 | | 203 | 11.276 | 11.025 | -0.251 |
| **KIG 367** | 5 | 15.28 | 13.08 | 11.64 | 7700 | 2 | | 35.29 | | | | |
| WISEA J094539.06+554811.5 | -5 | 16.49 | 13.07 | | 7691 | 4 | -9 | | 293 | 10.772 | 10.447 | -0.325 |
| **KIG 385** | 3 | 13.38 | 10.48 | 8.85 | 3084 | 4 | | 33.33 | | | | |
| SBS 0953+592 | 9 | 16.96 | | | 3083 | 1 | -1 | | 278 | 11.104 | 8.518 | -2.586 |
| **KIG 393** | 9 | 14.75 | 11.31 | 10.59 | 3065 | 68 | | 33.33 | | | | |
| CGCG 289-027 | -2 | 15.14 | 11.63 | | 2894 | 3 | -171 | | 239 | 10.408 | 12.916 | 2.508 |
| KUG 0958+599 | 8 | 16.44 | 14.39 | | 3416 | 1 | 351 | | 288 | 10.408 | 13.622 | 3.214 |
| **KIG 400** | 3 | 13.53 | 9.23 | 8.7 | 5087 | 8 | | 34.42 | | | | |
| KUG 1004+321B | 0 | 17.16 | | | 5157 | 4 | 70 | | 138 | 11.600 | 11.902 | 0.302 |
| KUG 1004+321A | 6 | 16.60 | | | 5248 | 1 | 161 | | 165 | 11.600 | 12.703 | 1.103 |
| **KIG 401** | 4 | 14.80 | 11.32 | 10.55 | 6214 | 7 | | 34.84 | | | | |
| SHOC 292 | 9 | 17.88 | | | 6144 | 2 | -70 | | 31 | 11.028 | 11.253 | 0.225 |
| **KIG 403** | 5 | 15.51 | 12.18 | 11.91 | 11628 | 5 | | 36.19 | | | | |
| WISEA J101400.65+355847.3 | 10 | 17.77 | 14.01 | | 11656 | 3 | 28 | | 327 | 11.024 | 11.480 | 0.456 |
| **KIG 416** | 5 | 14.79 | 11.67 | 10.16 | 2171 | 7 | | 32.94 | | | | |
| WISEA J102641.93+115350.9 | 9 | 17.67 | | | 2127 | 1 | -44 | | 189 | 10.424 | 11.635 | 1.211 |
| RFGC 1787 | 7 | 17.07 | 14.20 | | 2271 | 2 | 100 | | 203 | 10.424 | 12.378 | 1.954 |
| **KIG 423** | 2 | 13.90 | 9.77 | 8.94 | 6160 | 3 | | 34.78 | | | | |
| WISEA J103139.99+245046.1 | 4 | 16.79 | 13.92 | | 6133 | 3 | -27 | | 71 | 11.648 | 10.785 | -0.863 |

Table 1: (Contd.)

| Galaxy name | T | $b_t$ | $K_t$ | $K_B$ | $V_{LG}$ | $\sigma V_{LG}$ | $\Delta V_{LG}$ | modbest | $R_p$ | $\log(L_K)$ | $\log(M_{orb})$ | $\log M_{orb} - \log L_K$ |
|---|---|---|---|---|---|---|---|---|---|---|---|---|
| | | mag | mag | mag | km s$^{-1}$ | km s$^{-1}$ | km s$^{-1}$ | mag | kpc | L$_\odot$ | M$_\odot$ | |
| (1) | (2) | (3) | (4) | (5) | (6) | (7) | (8) | (9) | (10) | (11) | (12) | (13) |
| WISEA J103147.15+245128.1 | 8 | 17.99 | | | 6266 | 8 | 106 | | 105 | 11.648 | 12.143 | 0.495 |
| **KIG 444** | 6 | 13.09 | 10.23 | 9.81 | 4279 | 8 | | 34.05 | | | | |
| WISEA J105049.21-020636.1 | 9 | 18.86 | | | 4255 | 64 | -24 | | 66 | 11.008 | 10.653 | -0.355 |
| WISEA J105115.32-021402.6 | -5 | 18.98 | | | 4284 | 89 | 5 | | 130 | 11.008 | 9.580 | -1.428 |
| WISEA J105025.50-021627.2 | 0 | 17.72 | | | 4288 | 50 | 9 | | 218 | 11.008 | 10.318 | -0.690 |
| **KIG 467** | 0 | 14.49 | 10.20 | 10.33 | 6363 | 4 | | 34.90 | | | | |
| SDSS J110920.51+360537.5 | 9 | 18.36 | | | 6503 | 60 | 140 | | 121 | 11.140 | 12.447 | 1.307 |
| **KIG 471** | 2 | 15.32 | 11.84 | 10.68 | 8850 | 64 | | 35.60 | | | | |
| PGC1098042 | 9 | 17.39 | 13.84 | | 8849 | 64 | -1 | | 282 | 11.280 | 8.518 | -2.762 |
| **KIG 474** | 6 | 15.40 | | 11.82 | 5723 | 4 | | 34.68 | | | | |
| SDSS J111624.10+110611.5 | 5 | 17.91 | | | 5819 | 8 | 96 | | 130 | 10.456 | 12.149 | 1.693 |
| **KIG 476** | 6 | 15.28 | 13.33 | 10.93 | 3196 | 6 | | 33.46 | | | | |
| 2MASS J11204044+4301201 | 9 | 16.93 | | | 3022 | 4 | -174 | | 283 | 10.324 | 13.005 | 2.681 |
| **KIG 483** | 1 | 14.24 | 10.34 | 10.05 | 6457 | 7 | | 34.94 | | | | |
| SDSS J113205.82+033344.3 | 9 | 18.40 | 15.66 | | 6417 | 2 | -40 | | 218 | 11.268 | 11.614 | 0.346 |
| **KIG 488** | 3 | 15.82 | 10.90 | 11.09 | 12671 | 20 | | 36.36 | | | | |
| WISEA J113633.11+732811.2 | 3 | 16.84 | 11.80 | | 12595 | 116 | -76 | | 193 | 11.420 | 12.117 | 0.697 |
| **KIG 495** | 0 | 15.83 | 13.65 | 11.64 | 5365 | 2 | | 34.54 | | | | |
| WISEA J114056.77+012739.5 | 9 | 19.31 | 18.69 | | 5277 | 89 | -88 | | 55 | 10.472 | 11.701 | 1.229 |
| **KIG 497** | 3 | 15.17 | 11.95 | 11.03 | 7367 | 4 | | 35.22 | | | | |
| WISEA J114327.55+032738.1 | 9 | 17.28 | 14.41 | | 7833 | 2 | 466 | | 140 | 10.988 | 13.555 | 2.567 |
| **KIG 499** | 3 | 14.78 | 11.1 | 10.31 | 8374 | 11 | | 35.49 | | | | |
| WISEA J114804.59+014918.6 | 9 | 18.78 | | | 8550 | 64 | 176 | | 190 | 11.384 | 12.841 | 1.457 |
| **KIG 502** | 6 | 13.95 | 13.25 | 10.38 | 1529 | 3 | | 32.90 | | | | |
| SDSS J114810.61-015920.8 | 8 | 17.08 | | | 1351 | 2 | -178 | | 86 | 10.320 | 12.508 | 2.188 |
| **KIG 508** | 5 | 14.42 | 11.70 | 10.56 | 5952 | 2 | | 34.77 | | | | |
| PGC 1199172 | 7 | 18.88 | 15.96 | | 6123 | 89 | 171 | | 270 | 10.996 | 12.969 | 1.973 |

Table 1: (Contd.)

| Galaxy name | T | $b_t$ | $K_t$ | $K_B$ | $V_{LG}$ | $\sigma V_{LG}$ | $\Delta V_{LG}$ | modbest | $R_p$ | $\log(L_K)$ | $\log(M_{\text{orb}})$ | $\log M_{\text{orb}} - \log L_K$ |
|---|---|---|---|---|---|---|---|---|---|---|---|---|
| | | mag | mag | mag | km s$^{-1}$ | km s$^{-1}$ | km s$^{-1}$ | mag | kpc | L$_\odot$ | M$_\odot$ | |
| (1) | (2) | (3) | (4) | (5) | (6) | (7) | (8) | (9) | (10) | (11) | (12) | (13) |
| **KIG 512** | 4 | 13.48 | 11.24 | 9.7 | 1715 | 3 | | 32.25 | | | | |
| PGC 3291881 | 9 | 18.81 | 18.69 | | 1628 | 89 | -87 | | 240 | 10.332 | 12.330 | 1.998 |
| **KIG 516** | 4 | 15.21 | 12.23 | 10.69 | 6362 | 4 | | 34.91 | | | | |
| WISEA J115958.53+175344.1 | 8 | 17.66 | | | 6808 | 2 | 446 | | 313 | 11.000 | 13.866 | 2.866 |
| **KIG 520** | 3 | 14.30 | 10.72 | 10.28 | 7991 | 2 | | 35.37 | | | | |
| SDSS J121042.31+563228.1 | 8 | 18.25 | | | 8100 | 4 | 109 | | 262 | 11.348 | 12.565 | 1.217 |
| **KIG 525** | 3 | 13.80 | 10.46 | 9.68 | 6932 | 2 | | 35.08 | | | | |
| WISEA J122226.55+404541.7 | 8 | 19.84 | | | 6899 | 3 | -33 | | 305 | 11.472 | 11.593 | 0.058 |
| **KIG 528** | 5 | 13.44 | 9.62 | 9.33 | 4190 | 3 | | 34.01 | | | | |
| SDSSJ122429.57+483311.6 | 8 | 17.25 | | | 4069 | 8 | -121 | | 266 | 11.184 | 12.663 | 1.479 |
| **KIG 533** | -5 | 16.02 | 11.96 | 11.82 | 21510 | 27 | | 37.55 | | | | |
| WISEA J123143.94-010019.5 | 3 | 18.40 | 15.35 | | 21363 | 7 | -147 | | 30 | 11.852 | 11.884 | 0.032 |
| **KIG 539** | 3 | 13.98 | 10.27 | 9.82 | 5456 | 3 | | 34.55 | | | | |
| WISEA J123538.14+541350.1 | -2 | 17.69 | | | 5485 | 14 | 29 | | 75 | 11.204 | 10.869 | -0.335 |
| **KIG 540** | 8 | 14.34 | 11.77 | 11.26 | 2429 | 2 | | 32.91 | | | | |
| LCRS B123512.8-021530 | 9 | 17.35 | | | 2204 | 1 | -225 | | 194 | 9.972 | 13.064 | 3.092 |
| WISEA J123905.75-020044.4 | 8 | 16.74 | 14.48 | | 2459 | 1 | 30 | | 253 | 9.972 | 11.430 | 1.458 |
| **KIG 554** | 3 | 15.37 | 11.56 | 11.22 | 12999 | 5 | | 36.43 | | | | |
| SBS 1250+594 | 9 | 17.33 | | | 13311 | 2 | 312 | | 133 | 11.396 | 13.184 | 1.788 |
| **KIG 565** | 2 | 15.80 | 11.94 | 11.40 | 12110 | 4 | | 36.29 | | | | |
| PGC 3123173 | 9 | 18.15 | | | 12242 | 2 | 132 | | 315 | 11.268 | 12.811 | 1.543 |
| **KIG 589** | 1 | 15.80 | 11.94 | 11.48 | 18151 | 3 | | 37.17 | | | | |
| WISEA J133052.31+582119.3 | -2 | 17.83 | 13.76 | | 17885 | 3 | -266 | | 205 | 11.588 | 13.233 | 1.645 |
| **KIG 592** | 1 | 16.04 | 12.42 | 11.62 | 6579 | 5 | | 34.98 | | | | |
| WISEA J133156.94+030703.7 | 7 | 17.96 | 15.51 | | 6525 | 5 | -54 | | 229 | 10.656 | 11.896 | 1.240 |
| **KIG 604** | 1 | 12.17 | 8.38 | 7.66 | 1902 | 2 | | 32.33 | | | | |
| WISEA J135747.42+470115.6 | 9 | 18.22 | | | 2203 | 1 | 301 | | 188 | 11.180 | 13.303 | 2.123 |

**Table 1**: (Contd.)

| Galaxy name | T | $b_t$ | $K_t$ | $K_B$ | $V_{LG}$ | $\sigma V_{LG}$ | $\Delta V_{LG}$ | modbest | $R_p$ | $\log(L_K)$ | $\log(M_{\rm orb})$ | $\log M_{\rm orb} - \log L_K$ |
|---|---|---|---|---|---|---|---|---|---|---|---|---|
| | | mag | mag | mag | km s$^{-1}$ | km s$^{-1}$ | km s$^{-1}$ | mag | kpc | L$_\odot$ | M$_\odot$ | |
| (1) | (2) | (3) | (4) | (5) | (6) | (7) | (8) | (9) | (10) | (11) | (12) | (13) |
| **KIG 605** | 1 | 12.82 | 9.25 | 8.45 | 2407 | 3 | | 32.87 | | | | |
| SDSSJ135709.93+291310.4 | 9 | 17.45 | | | 2278 | 1 | -129 | | 53 | 11.080 | 12.017 | 0.937 |
| WISEA J135729.48+290332.7 | 7 | 19.3 | | | 2289 | 3 | -118 | | 115 | 11.080 | 12.276 | 1.196 |
| **KIG 606** | 3 | 15.20 | 12.24 | 10.75 | 14604 | 3 | | 36.69 | | | | |
| WISEA J135942.79+010637.1 | 4 | 16.39 | 12.53 | | 14484 | 1 | -120 | | 143 | 11.688 | 12.386 | 0.698 |
| **KIG 609** | 3 | 14.51 | 11.07 | 10.32 | 2002 | 2 | | 32.51 | | | | |
| SDSS J140003.25+461712.5 | 10 | 17.63 | | | 2225 | 3 | 223 | | 193 | 10.188 | 13.054 | 2.866 |
| SBS 1400+461 | 9 | 15.28 | 12.41 | | 2216 | 2 | 214 | | 264 | 10.188 | 13.154 | 2.966 |
| **KIG 612** | 3 | 15.03 | 11.47 | 10.97 | 8272 | 4 | | 35.46 | | | | |
| WISEA J140812.40+295050.3 | 6 | 18.09 | | | 8304 | 3 | 32 | | 141 | 11.108 | 11.230 | 0.122 |
| **KIG 618** | 2 | 15.97 | 12.35 | 11.45 | 11487 | 33 | | 36.17 | | | | |
| WISEA J141324.90+174658.9 | 2 | 17.75 | | | 11408 | 2 | -79 | | 88 | 11.200 | 11.812 | 0.612 |
| **KIG 625** | 4 | 14.60 | 10.59 | 9.57 | 4674 | 2 | | 34.24 | | | | |
| CGCG 133-37 | 5 | 15.89 | 13.43 | | 4589 | 2 | -85 | | 200 | 11.180 | 12.230 | 1.050 |
| **KIG 626** | 5 | 12.77 | 9.95 | 9.03 | 1576 | 6 | | 32.25 | | | | |
| LEDA 1150546 | 10 | 18.62 | 17.08 | | 1534 | 8 | -42 | | 247 | 10.600 | 11.711 | 1.111 |
| **KIG 630** | 3 | 14.14 | 10.59 | 9.83 | 4001 | 2 | | 33.91 | | | | |
| SDSS J142514.73+482727.4 | 7 | 17.62 | | | 4114 | 2 | 113 | | 201 | 10.944 | 12.480 | 1.536 |
| **KIG 634** | 5 | 12.02 | 8.29 | 8.2 | 2063 | 2 | | 32.24 | | | | |
| PGC 2573480 | 8 | 16.49 | | | 2006 | 2 | -57 | | 56 | 10.928 | 11.330 | 0.402 |
| **KIG 637** | 0 | 12.59 | 8.98 | 8.44 | 2267 | 23 | | 32.50 | | | | |
| PGC 2472512 | -5 | 17.37 | | | 2189 | 10 | -78 | | 38 | 10.936 | 11.436 | 0.500 |
| **KIG 638** | 5 | 12.48 | 8.98 | 8.1 | 1707 | 6 | | 31.58 | | | | |
| SDSS J143549.94+023618.4 | 7 | 16.80 | | | 1509 | 6 | -198 | | 290 | 10.704 | 13.127 | 2.423 |
| WISEA J143939.33+023454.5 | 8 | 17.22 | | | 1600 | 5 | -107 | | 296 | 10.704 | 12.602 | 1.898 |
| **KIG 642** | 7 | 14.45 | 12.39 | 10.8 | 1582 | 3 | | 32.85 | | | | |
| PGC 2043836 | 9 | 16.25 | | | 1582 | 2 | 0 | | 67 | 10.132 | 7.845 | -2.287 |

Table 1: (Contd.)

| Galaxy name | T | $b_t$ | $K_t$ | $K_B$ | $V_{LG}$ | $\sigma V_{LG}$ | $\Delta V_{LG}$ | modbest | $R_p$ | $\log(L_K)$ | $\log(M_{\mathrm{orb}})$ | $\log M_{\mathrm{orb}} - \log L_K$ |
|---|---|---|---|---|---|---|---|---|---|---|---|---|
| | | mag | mag | mag | km s$^{-1}$ | km s$^{-1}$ | km s$^{-1}$ | mag | kpc | L$_\odot$ | M$_\odot$ | |
| (1) | (2) | (3) | (4) | (5) | (6) | (7) | (8) | (9) | (10) | (11) | (12) | (13) |
| AGC 245249 | 10 | 17.3 | | | 1585 | 5 | 3 | | 272 | 10.132 | 9.462 | -0.670 |
| **KIG 644** | -2 | 15.21 | 10.94 | 10.55 | 8212 | 4 | | 35.43 | | | | |
| 2MASX J14440953+4327554 | -2 | 16.54 | 13.08 | | 8232 | 5 | 20 | | 255 | 11.264 | 11.079 | -0.185 |
| **KIG 653** | 1 | 13.71 | 9.55 | 8.94 | 4990 | 6 | | 34.68 | | | | |
| SDSSJ145055.4+403125.8 | -5 | 18.33 | | | 4911 | 2 | -79 | | 205 | 11.608 | 12.179 | 0.571 |
| **KIG 694** | 1 | 15.36 | 11.07 | 10.5 | 10606 | 2 | | 35.99 | | | | |
| PGC 1212133 | 3 | 17.63 | 14.05 | | 10628 | 4 | 22 | | 179 | 11.508 | 11.009 | -0.499 |
| **KIG 696** | 1 | 15.91 | 11.70 | 10.83 | 10039 | 4 | | 35.86 | | | | |
| WISEA J154223.60-011629.6 | -5 | 16.50 | 12.16 | | 9782 | 3 | -257 | | 154 | 11.324 | 13.079 | 1.755 |
| WISEA J154234.30-011934.1 | 5 | 18.25 | | | 9728 | 9 | -311 | | 326 | 11.324 | 13.570 | 2.246 |
| **KIG 712** | 4 | 12.87 | 9.52 | 8.94 | 1925 | 1 | | 32.38 | | | | |
| WISEA J155445.06+143503.1 | 8 | 18.30 | | | 2045 | 5 | 120 | | 72 | 10.688 | 12.086 | 1.398 |
| **KIG 721** | 1 | 15.26 | 11.97 | 11.02 | 11513 | 11 | | 36.16 | | | | |
| WISEA J160809.71+391430.0 | 10 | 18.84 | | | 11844 | 8 | 331 | | 239 | 11.368 | 13.490 | 2.122 |
| WISEA J160810.96+390811.6 | 3 | 17.53 | 14.04 | | 11464 | 3 | -49 | | 267 | 11.368 | 11.878 | 0.510 |
| WISEA J160755.71+390548.0 | 0 | 18.18 | 13.92 | | 11590 | 7 | -77 | | 287 | 11.368 | 12.301 | 0.933 |
| **KIG 738** | 1 | 15.88 | 11.97 | 11.49 | 10024 | 4 | | 35.85 | | | | |
| AGC 268204 | 3 | 16.99 | | | 10017 | 4 | -7 | | 259 | 11.056 | 10.176 | -0.880 |
| **KIG 739** | 4 | 15.81 | 12.43 | 11.41 | 8719 | 16 | | 35.55 | | | | |
| LEDA 1495610 | 9 | 16.89 | 13.89 | | 8803 | 3 | 84 | | 267 | 10.968 | 12.346 | 1.378 |
| **KIG 769** | 1 | 13.41 | 10.03 | 9.1 | 4414 | 2 | | 34.09 | | | | |
| WISEA J164340.85+223239.7 | 9 | 18.27 | | | 4395 | 1 | -19 | | 308 | 11.308 | 11.117 | -0.191 |
| **KIG 782** | -2 | 15.53 | 10.60 | 10.38 | 7430 | 21 | | 35.19 | | | | |
| PGC 1108385 | 0 | 16.42 | 11.50 | | 7637 | 38 | 207 | | 214 | 11.236 | 12.718 | 1.482 |
| **KIG 791** | 3 | 14.73 | 10.40 | 9.9 | 2683 | 13 | | 33.05 | | | | |
| KKR 33 | 8 | 16.70 | | | 2702 | 4 | 19 | | 327 | 10.572 | 11.143 | 0.571 |
| **KIG 807** | 4 | 15.53 | 11.65 | 11.47 | 8428 | 4 | | 35.47 | | | | |

**Table 1**: (Contd.)

| Galaxy name | T | $b_t$ | $K_t$ | $K_B$ | $V_{LG}$ | $\sigma V_{LG}$ | $\Delta V_{LG}$ | modbest | $R_p$ | $\log(L_K)$ | $\log(M_{\text{orb}})$ | $\log M_{\text{orb}} - \log L_K$ |
|---|---|---|---|---|---|---|---|---|---|---|---|---|
| | | mag | mag | mag | km s$^{-1}$ | km s$^{-1}$ | km s$^{-1}$ | mag | kpc | L$_\odot$ | M$_\odot$ | |
| (1) | (2) | (3) | (4) | (5) | (6) | (7) | (8) | (9) | (10) | (11) | (12) | (13) |
| PGC 3138681 | 8 | 18.07 | | | 8408 | 2 | -20 | | 324 | 10.912 | 11.185 | 0.273 |
| **KIG 813** | 5 | 12.38 | 9.48 | 8.75 | 1570 | 3 | | 31.86 | | | | |
| IC 4660 | 3 | 14.23 | 11.35 | | 1492 | 2 | -78 | | 208 | 10.556 | 12.173 | 1.617 |
| **KIG 853** | 8 | 15.64 | | 12.32 | 2592 | 5 | | 32.91 | | | | |
| $18^h 28^m 40^s + 32°02'50''$ | 10 | 18.0 | | | 2581 | 30 | -11 | | 159 | 9.548 | 10.356 | 0.808 |
| **KIG 882** | 4 | 15.62 | 11.38 | 10.70 | 6553 | 13 | | 34.87 | | | | |
| PGC 1104611 | 9 | 17.0 | 11.28 | | 6486 | 19 | -67 | | 120 | 10.980 | 11.803 | 0.823 |
| **KIG 884** | 4 | 15.67 | 11.23 | 10.42 | 8246 | 9 | | 35.38 | | | | |
| PGC 1172011 | 9 | 17.35 | | | 8188 | 3 | -58 | | 228 | 11.296 | 11.957 | 0.661 |
| **KIG 891** | 5 | 15.40 | 11.73 | 11.26 | 9308 | 45 | | 35.61 | | | | |
| PGC 1146821 | 7 | 17.40 | | | 9298 | 3 | -10 | | 221 | 11.052 | 10.415 | -0.637 |
| **KIG 892** | 4 | 14.74 | 10.69 | 10.53 | 9276 | 3 | | 35.63 | | | | |
| SDSSJ 205201.30+000535.8 | 9 | 19.62 | | | 9258 | 2 | -18 | | 199 | 11.312 | 10.881 | -0.431 |
| **KIG 938** | 8 | 15.56 | | 12.23 | 1967 | 2 | | 32.12 | | | | |
| AGC 310467 | 9 | 17.80 | | | 1983 | 3 | 16 | | 79 | 9.268 | 10.376 | 1.108 |
| PGC 1440937 | 0 | 17.26 | | | 1953 | 3 | -4 | | 216 | 9.268 | 9.613 | 0.345 |
| **KIG 949** | 7 | 14.60 | 13.08 | 10.69 | 1940 | 1 | | 31.78 | | | | |
| AGC 748687 | 10 | 18.7 | | | 1908 | 3 | -32 | | 289 | 9.748 | 11.543 | 1.795 |
| **KIG 952** | -2 | 15.76 | 10.89 | 11.45 | 7845 | 11 | | 35.25 | | | | |
| SDSSJ221018.28+163642.6 | 0 | 17.46 | 13.05 | | 7862 | 2 | 17 | | 260 | 10.832 | 10.949 | 0.117 |
| **KIG 962** | 4 | 15.54 | 11.81 | 10.76 | 7192 | 4 | | 35.07 | | | | |
| AGC 321476 | 9 | 18.0 | | | 7363 | 4 | 171 | | 167 | 11.036 | 12.760 | 1.724 |
| **KIG 964** | 2 | 15.66 | 11.51 | 11.29 | 7065 | 23 | | 35.02 | | | | |
| AGC 321503 | 9 | 18.2 | | | 7152 | 4 | 87 | | 238 | 10.804 | 12.326 | 1.522 |
| **KIG 990** | 9 | 15.76 | | 12.3 | 3656 | 2 | | 33.54 | | | | |
| AGC 748719 | 8 | 18.8 | | | 3566 | 4 | -90 | | 313 | 9.808 | 12.476 | 2.668 |
| **KIG 993** | 4 | 15.28 | 12.01 | 10.43 | 12926 | 9 | | 36.37 | | | | |

**Table 1**: (Contd.)

| Galaxy name | T | $b_t$ | $K_t$ | $K_B$ | $V_{LG}$ | $\sigma V_{LG}$ | $\Delta V_{LG}$ | modbest | $R_p$ | $\log(L_K)$ | $\log(M_{\rm orb})$ | $\log M_{\rm orb} - \log L_K$ |
|---|---|---|---|---|---|---|---|---|---|---|---|---|
| | | mag | mag | mag | km s$^{-1}$ | km s$^{-1}$ | km s$^{-1}$ | mag | kpc | L$_\odot$ | M$_\odot$ | |
| (1) | (2) | (3) | (4) | (5) | (6) | (7) | (8) | (9) | (10) | (11) | (12) | (13) |
| WISEA J225247.29+244439.7 | -2 | 18.16 | 13.52 | | 13000 | 4 | 74 | | 149 | 11.688 | 11.984 | 0.296 |
| **KIG 1001** | 1 | 13.76 | 9.63 | 9.34 | 3277 | 4 | | 33.27 | | | | |
| WISEA J225650.76-005032.8 | 7 | 18.34 | | | 3348 | 7 | 71 | | 179 | 10.884 | 12.025 | 1.141 |
| **KIG 1005** | 5 | 15.22 | 12.46 | 10.72 | 5129 | 2 | | 34.28 | | | | |
| MRK 1129 | 9 | 17.23 | 13.60 | | 5016 | 9 | -113 | | 133 | 10.736 | 12.301 | 1.565 |
| **KIG 1006** | 5 | 14.45 | 10.63 | 10.63 | 5753 | 16 | | 34.57 | | | | |
| AGC 335574 | 10 | 19.5 | | | 5639 | 6 | -114 | | 223 | 10.888 | 12.534 | 1.646 |
| **KIG 1019** | 4 | 13.41 | 9.70 | 9.16 | 3746 | 2 | | 33.59 | | | | |
| AGC 333285 | 8 | 18.8 | | | 3691 | 4 | -55 | | 245 | 11.084 | 11.942 | 0.858 |
| **KIG 1044** | 2 | 14.27 | 10.49 | 9.89 | 7123 | 5 | | 35.03 | | | | |
| AGC 333638 | 9 | 20 | | | 7183 | 4 | 60 | | 306 | 11.368 | 12.114 | 0.746 |